\documentclass[twocolumn,amsmath,amssymb]{revtex4}
\usepackage{graphicx}
\usepackage{bm}

\begin{document}
\newcommand{\uu}[1]{\underline{#1}}
\newcommand{\pp}[1]{\phantom{#1}}
\newcommand{\be}{\begin{eqnarray}}
\newcommand{\ee}{\end{eqnarray}}
\newcommand{\ve}{\varepsilon}
\newcommand{\vs}{\varsigma}
\newcommand{\Tr}{{\,\rm Tr\,}}
\newcommand{\pol}{{\textstyle\frac{1}{2}}}
\newcommand{\ba}{\begin{array}}
\newcommand{\ea}{\end{array}}
\newcommand{\bea}{\begin{eqnarray}}
\newcommand{\eea}{\end{eqnarray}}

\title{Quantum optics in different representations of the algebra of canonical commutation relations (I): Unexpected properties of Rabi oscillations --- theory and experiment}

\author{Marcin Wilczewski}%
\email{marcin@mif.pg.gda.pl}
\author{Marek Czachor}
\email{mczachor@pg.gda.pl}
\affiliation{%
Katedra Fizyki Teoretycznej i Metod Matematycznych\\
Politechnika Gda\'nska, Narutowicza 11/12, 80--952 Gda\'nsk,
Poland}
\begin{abstract}
We discuss the Jaynes-Cummings model in different representations of the algebra of canonical commutation relations. The first conclusion is that all the irreducible representations lead to equivalent physical predictions. However, the reducible representation recently introduced as a candidate for `QED without infinities' leads to new effects. We analyze from this perspective the experiments on Rabi oscillations performed by the Kastler-Brossel Laboratory group from Paris. Surprisingly, the results seem to support the reducible representation approach. We also discuss possibilities of more definitive tests of the new formalism.
\end{abstract}

\maketitle

\section{Introduction}

It is known that different representations of the same Lie algebra correspond to different types of physical systems. Parameters that characterize the representations have a meaning of quantum numbers (angular momentum for the rotation group, mass and spin for the Poincar\'e group, and so on). In principle, if we find a new representation of some physically meaningful group or algebra we should seriously consider the possibility that the representation corresponds to a physical system, perhaps yet unknown. This is how the anyons were predicted \cite{anyons}. 

In quantum optics the central role is played by the Lie algebra of canonical commutation relations (CCR)
\be
{[a(\bm k),a(\bm k')^*]}
&=& \delta(\bm k,\bm k')I(\bm k).\label{genCCR}
\ee
The element $I(\bm k)$ commutes with all the elements of the algebra and, by the Schur lemma, is proportional to the identity operator if the representation is irreducible. 
Fields with different boundary conditions correspond to different
$I(\bm k)$ and $\delta(\bm k,\bm k')$, and thus to different representations of CCR.

Still, the problem seems deeper than that. First of all, the theorem of von Neumann \cite{vN} states that there exists an infinite number of inequivalent irreducible representations of CCR for systems with infinite numbers of degrees of freedom (here corresponding to the infinite number of different wave vectors). It is not clear which representation to choose, even if one keeps boundary conditions fixed and restricts the analysis to free fields. Secondly, as stressed by Dirac in his last published paper \cite{bib:dirac2}, one should not ignore the physical potential inherent in reducible representations.  These are unavoidable in quantum physics since tensor products of irreducible representations are not themselves irreducible. Composite quantum systems are inherently reducible.

One of the representations that gained particular popularity in quantum optics employs the Hilbert space of infinitely many harmonic oscillators. To each frequency of the field there corresponds a separate oscillator. Physically the representation is rather pathological, just to mention the problem with the infinite energy of the ground state. Mathematically, the representation is also pathological in the sense of involving a non-separable Hilbert space typical of infinite tensor products \cite{vN2}, a fact rarely mentioned in the quantum optics literature.

It is relevant to mention in this context that recently, in a series of papers \cite{1,2,3,4} one of us investigated the possibility of electromagnetic field quantization in terms of certain reducible representations of CCR. The main feature of the proposed formalism was that there was no link between the number of frequencies allowed by field boundary conditions and the number $N$ of oscillators used in construction of the Hilbert space. In particular, it made perfect sense to speak of fields modelled by a finite number of oscillators even if the number of frequencies in the field was infinite. And vice versa, a monochromatic field could be modelled with any $N$. 

It is easy to understand why this is possible if one thinks of each oscillator as a wave packet involving all the possible frequencies. The parameter $N$ occurring in the representation has a status of a quantum number, and the vacuum is a Bose-Einstein condensate of $N$ oscillators at zero temperature.  The limit 
$N\to\infty$ plays a role of a correspondence principle mapping the new formalism into the standard one. What is interesting, when it comes to computing averages it turns out that the wave function of the ground state plays a role of a cut-off function regularizing integrals at ultraviolet and infrared regimes. The new formalism  thus automatically introduces many elements that are imposed in an ad hoc manner in the standard one.

However, if in reality $N$ is large but finite then quantum field theory we know nowadays is the 
$N=\infty$ approximation of a more fundamental, and hopefully more consistent, $N<\infty$ theory. The question is: Why the $N=\infty$ approximation works so well, and where to look for experimental manifestations of a finite $N$? 

In this paper we give partial answers to both parts of the question. We concentrate on the Jaynes-Cummings model where we solve Heisenberg equations of motion for the atomic inversion operator. A difference with respect to the usual treatment of the problem is that we begin with a representation-independent level. The first conclusion we find at that stage is that all the irreducible representations of (\ref{genCCR}) produce physically equivalent results. 

As the next step we choose the reducible representation parametrized by $N<\infty$ and compute evolution of atomic inversion with different vacuum, thermal, and coherent-state initial conditions. 

Here the situation changes. But first of all it has to be stressed that in the limit $N\to\infty$ we recover the standard formulas. This fact is of crucial importance for the whole approach since it plays a role of a {\it correspondence principle\/} and guarantees that the new theory may be regarded as a generalization of the one based on irreducible representations. 

For finite $N$ there are differences which we compare with experimental data of \cite{Haroche}. 
What is interesting, it seems that the data are more consistent with a finite $N$ than $N=\infty$ ($N=280$ for the nearly vacuum state at $T=0.8$~K, $N=420$ for the coherent state with $0.4$ and $0.85$ photons in average). The main reason why the experiment does not fully support the standard theory is that for finite $N$ one expects a faster relaxation of Rabi oscillation than what one might expect on the basis of $N=\infty$, and this is precisely what seems to happen in the experiment. 

The relaxation occurring for finite $N$ is not a decay but a beat and therefore waiting sufficiently long one should see a revival of the Rabi oscillation. We show that cavities with lifetimes of a few hundred $\mu$s in principle allow for observation of the revival.
Of particular interest are the maser and mazer experimental setups 
\cite{Walther1,Walther2,Walther3,Lamb}, but we leave it for a future work.

If our intuitions are correct and the finite $N$ representations are more physical than the limiting $N=\infty$ case, then many additional questions have yet to be answered. In the paper that accompanies the present one we address the issues of Lorentz and gauge covariance and test the formalism on another exactly solvable model: Quantum fields produced by a classical current.

The present paper is organized as follows. In Sec.~II we discuss the Jaynes-Cummings model at a representation independent level. In Sec. III we assume that the representation is irreducible and conclude that all such representations imply equivalent physics. In Sec. IV we switch to the reducible representation involving $N<\infty$ oscillators. Then, in Sec.~V, we write the explicit form of the inversion operator $R_3(t)$ in this representation and, in Sec.~VI,  compute atomic inversion with different initial conditions. In particular, we discuss in detail the limit $N\to\infty$ and its links to the law of large numbers and renormalized parameters. Some technicalities and remarks about the reducible representation are moved into Appendices.

\section{Representation independent formulation of the Jaynes-Cummings model}

The Jaynes-Cummings model \cite{JC,Allen} represents a two-level atom interacting with a single mode of electromagnetic field in a cavity. The cavity boundary conditions imply that the set of free-field momenta is discrete and it is convenient to work from the outset with the discrete notation
\be
[a_j,a^\ast_k]
=\delta_{jk} I_j.
\label{ccrrel}
\ee
Here $\delta_{jk}$ is the Kronecker delta, the asterisk denotes the adjoint in the sense of $^*$-algebras, and $I_j^*=I_j$. We assume there exists a free-field Hamiltonian $H_0$ satisfying
\be
{[a_j,H_0]}&=& \omega_j a_j,\\
{[a_j^*,H_0]}&=& -\omega_j a_j^*.
\ee
Note that $H_0$ cannot, in general, be given by $\sum_j\omega_j a_j^*a_j$; the latter works only if $I_j$ is an identity, which we do not assume at the present stage.

Let us now select a frequency $\omega_{j_0}=\omega$ and assume that only this frequency couples to the two-level system. The corresponding CCR operators will be indexed by $\omega$, that is: $a_{j_0}=a_\omega$, $a_{j_0}^*=a_\omega^*$, $I_{j_0}=I_\omega$. We also split $H_0$ into two parts: $H_0^\perp$ commuting with $a_\omega$ and $a_\omega^*$, and 
$H_0^\parallel=\omega N_\omega$, where
\be
{[a_\omega,N_\omega]}&=&  a_\omega,\\
{[a_\omega^*,N_\omega]}&=& - a_\omega^*.
\ee
Alternatively, we may assume that $N_\omega$ exists and define $H_0^\perp=H_0-\omega N_\omega$. 

The model is given by the full Hamiltonian
\be
H
=
\omega_0 R_3
+
H_0
+
g R_+ a_\omega
+
\bar g R_- a^\ast_\omega.
\ee
We employ the usual notation \cite{Allen} where $R_l=\sigma_l/2$, $R_\pm=R_1\pm i R_2$, 
$\sigma_l$ are the Pauli matrices, and $g$ is a complex coupling parameter.
It is useful to split $H$ into three mutually commuting parts: 
\be
H &=& H_0^\perp+\omega {\cal N}+\Omega,\\
{\cal N} &=& R_3+N_\omega,\\
\Omega
&=&
\Delta R_3
+
g R_+ a_\omega
+
\bar g R_- a^\ast_\omega.
\ee
$\Delta=\omega_0-\omega$ is the detuning. The evolution operator factorizes: 
\be
U_t
&=&
e^{-i H_0^\perp t}
e^{-i \omega {\cal N} t}
V_t,\label{U_t1}\\
V_t
&=&
e^{-i \Omega t}\label{U_t2}
=
\cos(\Omega_R t)-i
\frac{\sin(\Omega_R t)}{\Omega_R^2}
\Omega,\\
\Omega_R
&=&
\sqrt{\Delta^2/4+|g|^2 X},
\\
X
&=&
R_3 I_\omega
+
a_\omega^\ast a_\omega
+
\frac{1}{2}I_\omega.
\ee
Let us note that the operators ${\cal M}=R_3 I_\omega+a_\omega^\ast a_\omega$, occurring in $X$, and ${\cal N}=R_3+N_\omega$, occurring in $H$, should not be identified, although both are equivalent in irreducible representations of CCR, and both commute with $H$ independently of the choice of representation. Constructing $\cal M$ and $\cal N$ in the reducible representation we shall explicitly show that they are essentially different.

Only (\ref{U_t2}) is relevant in the context of Heisenberg-picture dynamics of $R_3(t)$ since 
$[R_3,H_0^\perp]=[R_3,{\cal N}]=0$. 
One can verify by a straightforward calculation that $id U_t/dt=U_t H$, and it is clear that the solution is valid for any representation of the Lie algebra (\ref{ccrrel}). 
Employing (\ref{U_t2}) one finds the Heisenberg picture evolution of $R_3$:
\begin{widetext}
\be
R_3(t)
=
R_3
\bigg(
1
-
2|g|^2  X
\frac{\sin^2({\Omega}_R t)}{{\Omega}^2_R}
\bigg)
+
\bigg(
\frac{\Delta}{2}
\frac{\sin^2({\Omega}_R t)}{{\Omega}^2_R}
-
i
\frac{\sin(2{\Omega}_R t)}{2{\Omega}_R}
\bigg)
g R_+ a_\omega
+
\bigg(
\frac{\Delta}{2}
\frac{\sin^2({\Omega}_R t)}{{\Omega}^2_R}
+
i
\frac{\sin(2{\Omega}_R t)}{2{\Omega}_R}
\bigg)
\bar g R_- a^*_\omega.\nonumber\\
&{}&
\label{R3tgen}
\ee
\end{widetext}
The proof of (\ref{R3tgen}) is outlined in the Appendix. 

The next important notion that can be introduced at a representation independent level is the displacement operator. 
The operator is defined in the usual way as
\be
D(z)
&=&
\exp\sum_j\big(z_j a^*_j-\bar z_j a_j\big)\label{D}
\ee
and satisfies 
\be
D(z)^* a_j D(z)
&=&
a_j+z_j I_j,\\
D(z)^* a_j^* D(z)
&=&
a^*_j+\bar z_j I_j,\\
D(z)^* I_j D(z)
&=&
I_j.
\ee
Acting with $D(z)$ on a vacuum vector we obtain a coherent state. Its form depends on what is meant by vacuum in a given representation.

\section{Evolution of atomic inversion operator in irreducible representations}

Assume we work in an irreducible representation with some carrier Hilbert space containing a vector $|0\rangle$ annihilated by all $a_\omega$. The abstract $*$-conjugation in the algebra can be replaced by Hermitian conjugation of operators in the representation. 
The representation satisfies 
\be
[a_j,a^\dag_k]={\cal Z} \delta_{jk} 1,
\label{ccri-ir}
\ee
where ${\cal Z}$ is a real positive number  (by Schur's lemma an element that commutes with all elements of the algebra is a constant times identity if the representation is irreducible, i.e. $I_j={\cal Z} 1$ for some ${\cal Z}$, and for all $j$; had ${\cal Z}$ been negative we would have called $a_j$ a {\it creation\/} operator, taken vacuum annihilated by $a^\dag_j$, and appropriately adjusted the notation). The solution 
(\ref{R3tgen}) involves the operator $|g|^2X$ whose representation reads
\be
|g|^2X
&=&
|g|^2 \big({\cal Z} R_3 
+
a_\omega^\dag a_\omega
+
{\cal Z}/2\big)\\
&=&
|\tilde g|^2 \big( R_3 
+
\tilde a_\omega^\dag \tilde a_\omega
+
1/2\big)=|\tilde g|^2 \tilde X,
\ee
where $\tilde a_\omega=a_\omega/\sqrt{{\cal Z}}$, 
$[\tilde a_j,\tilde a^\dag_k]=\delta_{jk} 1$, and $\tilde g=\sqrt{{\cal Z}}g$. 
Now, 
\begin{widetext}
\be
R_3(t)
=
R_3
\bigg(
1
-
2|\tilde g|^2  \tilde X
\frac{\sin^2({\tilde\Omega}_R t)}{{\tilde\Omega}^2_R}
\bigg)
+
\bigg(
\frac{\Delta}{2}
\frac{\sin^2({\tilde\Omega}_R t)}{{\tilde\Omega}^2_R}
-
i
\frac{\sin(2{\tilde\Omega}_R t)}{2{\tilde\Omega}_R}
\bigg)
\tilde g R_+ \tilde a_\omega
+
\bigg(
\frac{\Delta}{2}
\frac{\sin^2({\tilde\Omega}_R t)}{{\tilde\Omega}^2_R}
+
i
\frac{\sin(2{\tilde\Omega}_R t)}{2{\tilde\Omega}_R}
\bigg)
\bar{\tilde g} R_- \tilde a^\dag_\omega,\nonumber\\
&{}&
\label{R3t-ir}
\ee
\end{widetext}
where $\tilde\Omega=\sqrt{\Delta^2/4+|\tilde g|^2\tilde X}$. 
Clearly, the only difference between different representations is in the values of the coupling constant $\tilde g=g\sqrt{{\cal Z}}$. Since $g$ is proportional to the electron charge, different representations effectively differ by the value of the electron charge $\tilde e_0=e_0\sqrt{{\cal Z}}$. This type of rescaling is exactly what occurs in transition from the bare charge $e_0$ to the physical, renormalized charge $e_{\rm ph}=e_0\sqrt{Z_3}$ typical of renormalized electromagnetic fields. 
Let us note finally that 
\be
\tilde a_j D(z)|0\rangle
=
z_j\sqrt{\cal Z}D(z)|0\rangle \label{Z-eigen}
\ee
and thus the relation between the parameter $z$ and the `physical' amplitude $\tilde z$ is also renormalized: $\tilde z=z\sqrt{\cal Z}$. After these rescalings of the `bare' parameters we will obtain identical formulas for coherent state averages of 
$R_3(t)$, independently of our choice of the irreducible representation. We conclude that the theorem of von Neumann does not bring to our problem anything physically important. 

So let us switch to reducible representations.

\section{$N<\infty$ reducible representation}

The representation is constructed as follows. For simplicity we ignore the polarization degree of freedom. Take an operator $a$ satisfying $[a,a^{\dag}]=1$ and the kets $|\bm k\rangle$ corresponding to standing waves in some cavity. We define 
\be
a(\bm k) = |\bm k\rangle\langle \bm k|\otimes a,\quad
I(\bm k) = |\bm k\rangle\langle \bm k|\otimes 1.\label{I N=1}
\ee
The operators (\ref{I N=1}) satisfy (\ref{genCCR}), where
$\delta(\bm k,\bm k')$ is the 3D Kronecker delta. The fact that $I(\bm k)$ is not proportional to the identity means that the representation is reducible. In our terminology this is the `$N=1$ representation'. Its Hilbert space $\cal H$ is spanned by the kets 
$|\bm k,n\rangle=|\bm k\rangle|n\rangle$, where $a^{\dag}a|n\rangle=n|n\rangle$. Such a Hilbert space represents essentially a single harmonic oscillator of indefinite frequency \cite{1}. An important property of the representation is that  $\sum_k I(\bm k)=I$ is the identity operator in $\cal H$. 
A vacuum of this representation is given by any state annihilated by all $a(\bm k)$. The vacuum state is not unique and belongs to the subspace spanned by $|\bm k,0\rangle$. In our notation a $N=1$ vacuum state reads $|O\rangle=\sum_k O(\bm k)|\bm k,0\rangle$ and is normalized by $\sum_k |O(\bm k)|^2=\sum_k Z(\bm k)=1$. Such a vacuum is exactly analogous to a single-oscillator ground-state center-of-mass {\it wavepacket\/}. As shown in 
\cite{2,3} (for a more complete discussion see the companion paper \cite{4}) in a fully relativistic formulation the maximal probability $Z=\max_k\{Z(\bm k)\}$ is a
Poincar\'e invariant, an aspect worth keeping in mind. 
For $N\geq 1$ the representation space is given by the tensor power 
$\uu{\cal H}={\cal H}^{\otimes N}$, i.e. we take the Hilbert space of $N$ (bosonic) harmonic oscillators. Let $A: {\cal H}\to {\cal H}$ be any operator for $N=1$. 
We denote
$A^{(n)}=I^{\otimes (n-1)}\otimes A \otimes I^{\otimes (N-n)}$, 
$A^{(n)}: \uu{\cal H}\to \uu{\cal H}$, for $1\leq n\leq N$. For arbitrary $N$ the representation is defined by
\be
\uu a(\bm k)
&=&\frac{1}{\sqrt{N}}\sum_{n=1}^N a(\bm k)^{(n)},\label{aN} 
\\
\uu I(\bm k)&=&\frac{1}{N}\sum_{n=1}^N I(\bm k)^{(n)}, \label{IN}\\
{[\uu a(\bm k),\uu a(\bm k')^{\dag}]} &=& \delta(\bm k,\bm k')\uu I(\bm k),
\\
\sum_k \uu I(\bm k)&=&\uu I=I^{\otimes N}
\ee
and the $N$-oscillator vacuum is the $N$-fold tensor power of the $N=1$ case, a kind of Bose-Einstein condensate consisting of $N$ wavepackets:
\be
|\uu O\rangle=|O\rangle\otimes\dots\otimes|O\rangle=|O\rangle^{\otimes N}.\label{vacN}
\ee
The free-field Hamiltonian is, for $N=1$ and $\omega_k=|\bm k|$, 
\be
\textstyle
H_0=\sum_k \omega_ka(\bm k)^{\dag}a(\bm k)
=
\sum_k \omega_k|\bm k\rangle\langle \bm k|\otimes a^{\dag}a.
\ee
In each eigensubspace with fixed $|\bm k\rangle$ the operator $H_0$ is just an ordinary Hamiltonian of the oscillator with frequency $\omega_k$. For arbitrary $N$ the generator of free field evolution is the Hamiltonian of $N$ noninteracting oscillators, i.e. 
$\uu H_0=\sum_{n=1}^N H_0^{(n)}$.
Let us stress that $\uu H_0$ should not be confused with 
$\sum_k \omega_k\uu a(\bm k)^{\dag}\uu a(\bm k)$ which does not have a clear interpretation in this context and does not describe noninteracting oscillators. We have seen already that the operator $\uu a(\bm k)^{\dag}\uu a(\bm k)$ will nevertheless play an important role in the Jaynes-Cummings problem. Our definition of $\uu H_0$ implies that $[\uu a(\bm k),\uu H_0]=\omega_k\uu a(\bm k)$ which is the formula we required at the representation independent level. Let us now select some $\bm p$ with frequency $\omega=|\bm p|$ to be the mode that interacts with the two-level atom. To simplify further calculations we denote $\uu a(\bm p)=\uu a_\omega$, 
$\uu I(\bm p)=\uu I_\omega$, $|\bm p\rangle\langle\bm p|\otimes 1=P_1$, $P_0=I-P_1$. 

The projector $P_1$ is related to 
$\uu I_\omega$ by 
\be
\uu I_\omega
&=&
\frac{1}{N}\Big(P_1\otimes I\dots\otimes I+\dots+
I\otimes\dots\otimes I\otimes P_1\Big).
\label{I1}
\ee
One recognizes in (\ref{I1}) the frequency-of-success operator discussed in the context of quantum laws of large numbers \cite{Hartle,Farhi,Gutman,Casinello,Aharonov,Finkelstein,Caves}. Eigenvalues 
of $\uu I_\omega$ coincide with all the possible frequences of `heads' in $N$ trials of coin tossing, i.e. 
$\uu I_\omega=\sum_{s=0}^N(s/N)P(s/N)$ with spectral projectors
\be
P(s/N)
&=&
\sum_{s_1+\ldots+s_N=s; s_j=0,1}
P_{s_1}\otimes \ldots \otimes P_{s_N}.\label{P(s/N)}
\ee
The explicit form (\ref{P(s/N)}) shows that $P(s/N)$ commute with all $\uu a(\bm k)$, 
$\uu a(\bm k)^{\dag}$, and $\uu I(\bm k)$. Of particular importance is the splitting of the Jaynes-Cummings problem into subspaces corresponding to a given $s/N$. We define
\be
a(s) &=& \uu a_\omega P(s/N),\\
a(s)^{\dag} &=& \uu a_\omega^{\dag} P(s/N),
\ee
and obtain the representation 
\be
{[a(s),a(s)^{\dag}]} &=& (s/N) P(s/N)\label{ccr-s}
\ee
of CCR in the Hilbert space ${\cal H}(s)=P(s/N)\uu{\cal H}$. 

Each ${\cal H}(s)$ is an invariant subspace for the Jaynes-Cummings dynamics, and in each such a subspace we effectively deal with a representation given by (\ref{ccri-ir}) whose 
${\cal Z}=s/N$, as we shall see in the next Section.

\section{Evolution of atomic inversion operator in $N<\infty$ representation}

As the general formula (\ref{R3tgen}) is valid also in this representation and all the operators that occur there commute with $P(s/N)$, we begin with splitting $R_3(t)$ into parts acting in the invariant subspaces ${\cal H}(s)$. 
We employ the usual notation $R_3=(|+\rangle\langle+|-|-\rangle\langle-|)/2$ and treat 
$|+\rangle\langle+|+|-\rangle\langle-|$ as the identity operator (to be more exact we should tensor the atomic-space identity with the field-space identity, but we prefer this simplified convention). 

Denoting 
$X(s)=XP(s/N)$ we get $X=\sum_{s=0}^NX(s)$ and
\be
X(s)
&=&
\Big(\frac{s}{N}R_3+a(s)^{\dag}a(s)+\frac{s}{2N}\Big)P(s/N)\\
&=&
\frac{s}{N}|+\rangle\langle+|P(s/N)+a(s)^{\dag}a(s).
\ee
Commutation relation (\ref{ccr-s}) implies that eigenvalues of $a(s)^{\dag}a(s)$ are $sn/N$, 
$n=0,1,2,\dots$. In spectral representation 
\be
a(s)^{\dag}a(s)=\frac{s}{N}\sum_{n=0}^\infty n\Pi(n,s).
\ee
The spectral projectors satisfy $\Pi(n,s)P(s/N)=\Pi(n,s)$, 
$\sum_{n=0}^\infty \Pi(n,s)=P(s/N)$. Spectral representation of $X(s)$ therefore reads
\be
X(s)
&=&
\frac{s}{N}\sum_{n=0}^\infty
\big(|+\rangle\langle+|+n\big)\Pi(n,s)\\
&=&
\frac{s}{N}\sum_{n=0}^\infty n\hat\Pi(n,s).
\ee
Spectral projectors $\hat\Pi(n,s)$ of $X(s)$ are related to $\Pi(n,s)$ of $a(s)^{\dag}a(s)$ 
by 
\be
\hat\Pi(n,s)
&=&
|+\rangle\langle+|\Pi(n-1,s)+|-\rangle\langle-|\Pi(n,s),
\ee
for $n>0$, and $\hat\Pi(0,s)=|-\rangle\langle-|\Pi(0,s)$. 
In the Appendix we show that
$[R_+ a(s),\hat\Pi(n,s)]=[R_- a(s)^{\dag},\hat\Pi(n,s)]$. Evolution of the atomic inversion operator is given in our reducible representation by
\begin{widetext}
\be
R_3(t)
&=&
R_3
-
\sum_{s=0}^N
\sum_{n=0}^\infty
2 R_3|g|^2  \frac{ns}{N}
\frac{\sin^2\big(t\sqrt{\Delta^2/4+|g|^2 ns/N} \big)}
{\Delta^2/4+|g|^2 ns/N}
\hat\Pi(n,s)\nonumber\\
&\pp=&
\pp{R_3}
+
\sum_{s=0}^N
\sum_{n=0}^\infty
\bigg(
\frac{\Delta}{2}
\frac{\sin^2\big(t\sqrt{\Delta^2/4+|g|^2 ns/N} \big)}
{\Delta^2/4+|g|^2 ns/N}
-
i
\frac{\sin\big(2t\sqrt{\Delta^2/4+|g|^2 ns/N} \big)}
{2\sqrt{\Delta^2/4+|g|^2 ns/N}}
\bigg)
\hat\Pi(n,s)
g R_+ a(s)\nonumber\\
&\pp=&
\pp{R_3}
+
\sum_{s=0}^N
\sum_{n=0}^\infty
\bigg(
\frac{\Delta}{2}
\frac{\sin^2\big(t\sqrt{\Delta^2/4+|g|^2 ns/N} \big)}
{\Delta^2/4+|g|^2 ns/N}
+
i
\frac{\sin \big(2t\sqrt{\Delta^2/4+|g|^2 ns/N} \big)}
{2\sqrt{\Delta^2/4+|g|^2 ns/N}}
\bigg)
\bar g R_- a(s)^{\dag}\hat\Pi(n,s)\label{R3t-red}.
\ee
\end{widetext}
Formula (\ref{R3t-red}) can be directly compared to (\ref{R3t-ir}) by means of the following simple rule: Skip the sum over $s$ and set $s/N=1$. An alternative recipe is to skip the sum over $s$ and set $s/N={\cal Z}$, which corresponds to CCR with ${\cal Z}1$ at the right-hand-side. 
Yet another intuitive rule can be found by means of the law of large numbers and works for the weak limit $N\to\infty$. To make it precise one has to switch to the level of averages. 
We will see that the weak law of large numbers plays a role of a correspondence principle between our $N<\infty$ formalism and the standard regularized one.

\section{Evolution of atomic inversion in $N<\infty$ representation}

Acting on the vacuum state (\ref{vacN}) with the displacement operator (\ref{D}) one obtains a coherent state. Here we are interested in a monochromatic coherent state with frequency $\omega$
\be
|\uu z\rangle
&=&
\exp\big(z \uu a_\omega^{\dag}-\bar z \uu a_\omega\big)|\uu O\rangle\label{|beta>}.
\ee
The coherent state is not an eigenstate of the annihilation operator $\uu a_\omega$ but a direct sum of its eigenstates. Indeed, the decomposition
\be
|\uu z\rangle
&=&
\sum_{s=0}^N P(s/N)|\uu z\rangle
=
\sum_{s=0}^N |z(s)\rangle
\ee
accompanying $\uu a_\omega=\sum_{s=0}^N \uu a_\omega P(s/N)=\sum_{s=0}^N a(s)$ implies that 
\be
a(s)|z(s)\rangle=(s/N)z |z(s)\rangle.
\ee
The analogy of the latter eigenvalue problem to (\ref{Z-eigen}) is evident. 
Alternatively, one can say that the coherent state is a generalized eigenvector of 
$\uu a_\omega$, i.e.
\be
\uu a_\omega|\uu z\rangle
&=&
z \uu I_\omega 
|\uu z\rangle.
\ee
Another state of interest, particularly in the context of experiments, is the mixture
\be
\rho
&=&
\big(p_+
|+\rangle\langle+|
+
p_-
|-\rangle\langle-|
\big)
\rho_{\rm field}.\label{rho-f}
\ee

\subsection{Reduced inversion operator: Coherent state}

Evaluating an average of (\ref{R3t-red}) in an arbitrary coherent state 
$|\uu z\rangle$ we obtain the reduced operator 
\be
R_z(t)
=
\langle \uu z|R_3(t)|\uu z\rangle
\ee
involving only the atomic degrees of freedom:
\begin{widetext}
\be
R_z(t)
&=&
R_3
+
\sum_{s=0}^N
\frac{s}{N}
\sum_{n=0}^\infty
\langle \uu z|\Pi(n,s)|\uu z\rangle
\nonumber\\
&\pp=&
\pp{R_3}
\times
\Bigg(
-|g|^2  (n+1)
\frac{\sin^2\big(t\sqrt{\Delta^2/4+|g|^2 (n+1)s/N} \big)}
{\Delta^2/4+|g|^2 (n+1)s/N}
|+\rangle\langle+|
+
|g|^2  n
\frac{\sin^2\big(t\sqrt{\Delta^2/4+|g|^2 ns/N} \big)}
{\Delta^2/4+|g|^2 ns/N}
|-\rangle\langle-|
\nonumber\\
&\pp=&
\pp{R_3}
+
zg
\bigg(
\frac{\Delta}{2}
\frac{\sin^2\big(t\sqrt{\Delta^2/4+|g|^2 (n+1)s/N} \big)}
{\Delta^2/4+|g|^2 (n+1)s/N}
-
i
\frac{\sin\big(2t\sqrt{\Delta^2/4+|g|^2 (n+1)s/N} \big)}
{2\sqrt{\Delta^2/4+|g|^2 (n+1)s/N}}
\bigg)
R_+ \nonumber\\
&\pp=&
\pp{R_3}
+
\bar z\bar g 
\bigg(
\frac{\Delta}{2}
\frac{\sin^2\big(t\sqrt{\Delta^2/4+|g|^2 (n+1)s/N} \big)}
{\Delta^2/4+|g|^2 (n+1)s/N}
+
i
\frac{\sin \big(2t\sqrt{\Delta^2/4+|g|^2 (n+1)s/N} \big)}
{2\sqrt{\Delta^2/4+|g|^2 (n+1)s/N}}
\bigg)
R_- \Bigg),\label{R_z}
\ee
\end{widetext}
where (see Appendix)
\be
{}&{}&
\langle \uu z|\Pi(n,s)|\uu z\rangle
\label{<z|Pi|z>}
\\
&{}&\pp=
=
\frac{|z\sqrt{s/N}|^{2n}}{n!}
e^{-|z\sqrt{s/N}|^2}
\left(
\begin{array}{c}
N\\
s
\end{array}
\right)
Z_\omega^s (1-Z_\omega)^{N-s}.\nonumber
\ee
Here 
\be
Z_\omega=\langle \uu O|\uu I_\omega|\uu O\rangle=\langle O|I_\omega|O\rangle=|O(\bm p)|^2
\ee
is the probability of finding the momentum $\bm p$ corresponding to the resonant mode 
$\omega=|\bm p|$, if the vacuum state of the field is $|\uu O\rangle$.
In the last term of (\ref{<z|Pi|z>}) one recognizes the binomial distribution for $N$ Bernoulli trials with probability of success equal to $Z_\omega$.

\subsection{Vacuum-state initial condition: $|\uu z\rangle=|\uu O\rangle$}

Assume initially there are no photons and the atom is in either ground or excited state, i.e.
\be
|{\Psi}\rangle
=
|\pm\rangle|\underline{O}\rangle .
\ee
In this case only the $n=0$ term counts in (\ref{R_z}). The case of atomic ground state, 
$|{\Psi}\rangle=|-\rangle|\underline{O}\rangle$, is trivial since 
\be
w(t)=\langle {\Psi}|R_3(t) |{\Psi}\rangle=
\langle -|R_z(t) |-\rangle
=-1/2.
\ee
However, starting with the excited state
$|{\Psi}\rangle=|+\rangle|\underline{O}\rangle$ we find 
\begin{widetext}
\be
w(t)
=
\langle {\Psi}|R_3(t) |{\Psi}\rangle
=
\langle +|R_z(t) |+\rangle
=
\frac{1}{2}
-
\sum_{s=0}^N
|g|^2\frac{s}{N}
\frac{\sin^2\sqrt{\Delta^2/4+|g|^2s/N }t}{\Delta^2/4+|g|^2s/N}
\left(
\begin{array}{c}
N\\s
\end{array}
\right)
Z_\omega^s (1-Z_\omega)^{N-s}.
\label{R3t_1}
\ee
\end{widetext}
The law of large numbers for the binomial distribution implies that
\be
\lim_{N\to\infty}w(t)
=
\frac{1}{2}
-
|g|^2Z_\omega
\frac{\sin^2\sqrt{\Delta^2/4+|g|^2 Z_\omega }t}{\Delta^2/4+|g|^2 Z_\omega},
\label{R3t_1->}
\ee
i.e. the frequency $s/N$ approaches the probability of success in a single trial, that is 
$s/N\to Z_\omega$.

It is instructive to compare this result with the one we would have found had we started with the general irreducible representation whose right-hand-side is ${\cal Z}1$. The corresponding result reads
\be
w(t)
=
\frac{1}{2}
-
|g|^2 {\cal Z}
\frac{\sin^2\sqrt{\Delta^2/4+|g|^2 {\cal Z} }t}{\Delta^2/4+|g|^2 {\cal Z}}.
\label{R3t_1-Z}
\ee
These two formulas appear so similar that one can easily overlook an important difference: The parameter $\cal Z$ in (\ref{R3t_1-Z}) is a number independent of $\omega$, whereas $Z_\omega$ in (\ref{R3t_1->}) is a value of a function vanishing for $\omega\to\infty$. Charge renormalization cannot be performed in the same way in both models, because charge must remain  a relativistic invariant. 

In the irreducible case the recipe is simple: $e_{\rm ph}=e_0 \sqrt{\cal Z}$. In the reducible case one first extracts from $Z_\omega$ the relativistic invariant $Z=\max_k\{Z(\bm k)\}$ \cite{2,3,4}. Then one writes $Z_\omega=Z\chi_\omega$, redefines charge
$e_{\rm ph}=e_0 \sqrt{Z}$, and finally 
\be
\lim_{N\to\infty}w(t)
=
\frac{1}{2}
-
|g_{\rm ph}|^2\chi_\omega
\frac{\sin^2\sqrt{\Delta^2/4+|g_{\rm ph}|^2 \chi_\omega }t}
{\Delta^2/4+|g_{\rm ph}|^2 \chi_\omega}.
\label{R3t_1->>}
\ee
The solutions (\ref{R3t_1-Z}) and (\ref{R3t_1->>}) are identical up to the presence of the cut-off function $\chi_\omega$ in (\ref{R3t_1->>}). 
Needless to say the function would necessarily regularize the interaction if we decided to work with extremely high frequencies $\omega$. 

Let us note that we did not introduce any cut-off in the Hamiltonian. The cut-off has appeared automatically through the structure of the vacuum state typical of the reducible representation. The theory gets regularized even though we do not really need it in such a simple example. The regularization neither solves here any problem nor spoils anything. If we assume that for optical frequencies $\chi_\omega\approx 1$ we obtain exact agreement between the irreducible case and the $N\to\infty$ limit of the reducible one. This is an example of the correspondence principle we have mentioned in the introduction. 

The next step is to understand if one really needs $N=\infty$, and if some finite $N$ cannot, in fact, be consistent with experimental data. We first rewrite (\ref{R3t_1}) by means of the renormalized coupling
\begin{widetext}
\be
w(t)
=
\frac{1}{2}
-
\sum_{s=0}^N
\frac{|g_{\rm ph}|^2}{Z}\frac{s}{N}
\frac{\sin^2\sqrt{\Delta^2/4+|g_{\rm ph}|^2s/(ZN) }t}{\Delta^2/4+|g_{\rm ph}|^2s/(ZN)}
\left(
\begin{array}{c}
N\\s
\end{array}
\right)
Z_\omega^s (1-Z_\omega)^{N-s}.
\ee
\end{widetext}
For $N$ large enough the binomial distribution can be approximated by a Gaussian
\be
\left(
\begin{array}{c}
N\\s
\end{array}
\right)
Z_\omega^s (1-Z_\omega)^{N-s}
\approx
\frac{e^{-\frac{(s-NZ)^2}{2NZ(1-Z)}}}{\sqrt{2\pi NZ(1-Z)}}
\approx
\frac{e^{-\frac{(s-NZ)^2}{2NZ}}}{\sqrt{2\pi NZ}}
\nonumber
\ee
whose shape is controlled mainly by the product $NZ$. The smaller $Z$ the less important its exact value (the second approximate equality holds for small $Z$). 
So $NZ$ is the parameter we may have a chance of seeing in experiments. Let us stress that both $N$ and $Z$ are relativistically invariant. 

The atomic inversion $w(t)$ is here a sum of $N$ oscillations, each at a
different Rabi frequency. 
It is clear that for finite $N$ the evolution of $w(t)$ will reveal collapses and revivals, and not a simple Rabi oscillation as would be expected on the basis of irreducible representations (cf. the analysis of the experiment in the next subsection, and in particular Fig.~6). The solution is similar to those known from the standard analysis of collapses and revivals of coherent-state evolutions \cite{Narozhny}.

\subsection{Experiment: Thermal mixture as the initial condition}

Replacing the Poisson statistics in (\ref{<z|Pi|z>}) by thermal probability we obtain a formula applicable to situations where the coherent light is replaced by a thermal mixture. We 
thus assume (\ref{rho-f}) with
\be
\Tr\rho_{\rm field}\Pi(n,s)
&=&
{\tt P}(n)
\left(
\begin{array}{c}
N\\s
\end{array}
\right)
Z_\omega^s (1-Z_\omega)^{N-s},\\
{\tt P}(n)
&=&
\frac{\bar n^n}{(1+\bar n)^{(n+1)}},
\ee
and find the formula directly applicable to experiments:
\begin{widetext}
\be
w(t)
&=&
\frac{p_+-p_-}{2}
+
\sum_{s=0}^N
\left(
\begin{array}{c}
N\\s
\end{array}
\right)
Z_\omega^s (1-Z_\omega)^{N-s}
\sum_{n=0}^\infty
{\tt P}(n)
\Big(\frac{p_-\bar n}{1+\bar n}-p_+\Big)
|g|^2\frac{(n+1)s}{N}
\frac{\sin^2\big(t\sqrt{\frac{\Delta^2}{4}+|g|^2 \frac{(n+1)s}{N}} \big)}
{\Delta^2/4+|g|^2 (n+1)s/N}.
\ee
Taking the limit $N\to\infty$ we obtain
\be
w(t)
&=&
\frac{p_+-p_-}{2}
+
\sum_{n=0}^\infty
{\tt P}(n)
\Big(\frac{p_-\bar n}{1+\bar n}-p_+\Big)
|g_{\rm ph}|^2\chi_\omega
\frac{\sin^2\big(t\sqrt{\Delta^2/4+|g_{\rm ph}|^2 (n+1)\chi_\omega} \big)}
{\Delta^2/4+|g_{\rm ph}|^2(n+1)\chi_\omega},
\ee
\end{widetext}
which, up to $\chi_\omega$, is known from irreducible representations.

In what follows we compare theoretical results with the precise data 
on optical Rabi oscillations reported by the Kastler-Brossel Laboratory group from Paris 
\cite{Haroche}. 
\begin{figure}
\includegraphics{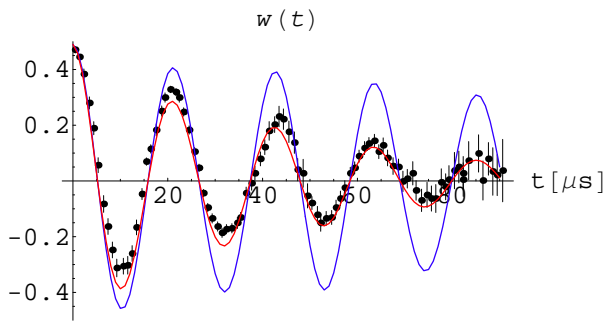}
    \caption{Data from the experiment vs. standard theoretical predictions. The blue line is the Rabi oscillation in cavity with photon lifetime $T_{\rm cav}=220$ $\mu$s. The red curve fits better but corresponds to $T_{\rm cav}=45$ $\mu$s. 
    }
\end{figure}

\begin{figure}
\includegraphics{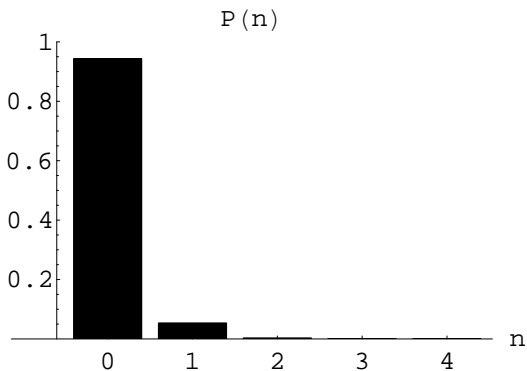}
    \caption{Thermal probability distribution with $\bar n=0.05$
    }
\end{figure}
\begin{figure}
\includegraphics{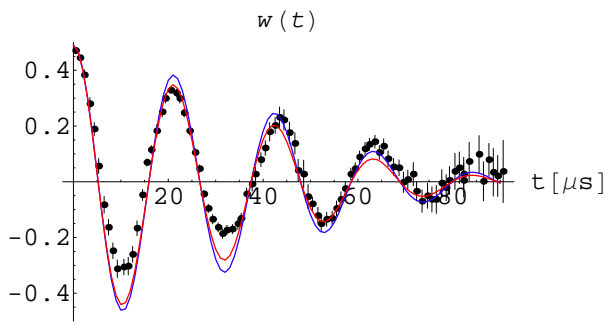}
    \caption{
  Data from the experiment vs. predictions based on the reducible representation with 
  $N Z=28$. The blue curve corresponds to an ideal cavity. The relaxation is not a decay but a beat: Waiting sufficiently long we will see a revival. The red line is the Rabi oscillation with additional damping corresponding to  $T_{\rm cav}=220$ $\mu$s.
      }
\end{figure}

\begin{figure}
\includegraphics{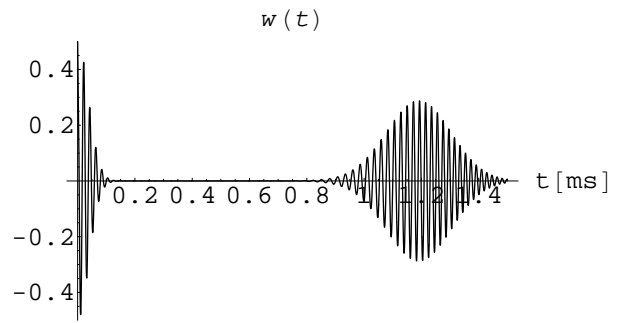}
    \caption{The same parameters as in Fig.~3 but for longer times and in an ideal cavity. 
    }
\end{figure}
\begin{figure}
\includegraphics{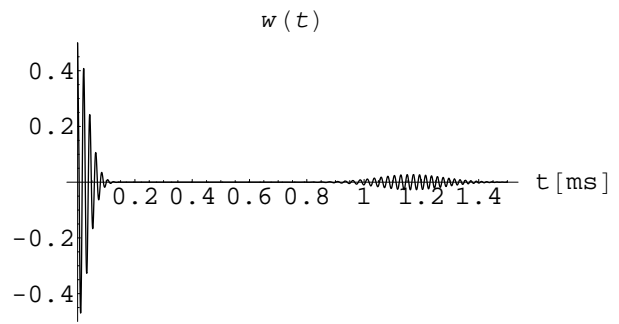}
    \caption{The same parameters as in Fig.~3 and Fig.~4 but additionally damped by 
    $\exp(-t/T_{\rm cav})$, with $T_{\rm cav}=500$ $\mu$s.
    }
\end{figure}
\begin{figure}
\includegraphics{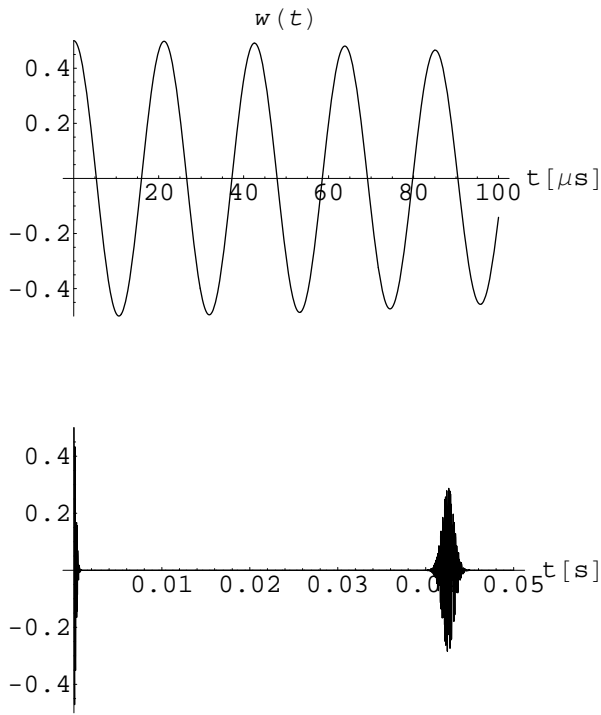}
    \caption{
    The same $w(t)$ at two different time scales, for $NZ=1000$, $g_{\rm ph}=47$~kHz, ideal cavity, and ideal vacuum. For $0<t<100$~$\mu$s (duration of the measurement reported in \cite{Haroche}) the plot is indistinguishable from the ideal undamped Rabi oscillation predicted by irreducible representations. For $0<t<0.05$~s we observe collapse and revival. The upper plot illustrates the idea behind the correspondence principle: For any finite time interval one can choose $NZ$ in a way guaranteeing an agreement, within some given error bars, with the standard theory. The lower plot shows that experiments involving sufficiently long times are in principle capable of discriminating between finite and infinite $N$s.
    }
\end{figure}
The relevant plot is Fig.~2A in \cite{Haroche}.  In Fig.~1 we show the data vs. standard theoretical predictions based on irreducible representations. We assume exact resonance condition $\Delta=0$, $g_{\rm ph}=47$ kHz, and $p_+=0.99$. As the initial field state we take the thermal mixture ${\tt P}(n)$ for $\bar n=0.05$ 
(Fig.~2 shows that ${\tt P}(n)$ agrees with Fig.~2$\alpha$ from \cite{Haroche}). 
The blue curve represents the Rabi oscillation additionally damped by the factor 
$\exp(-t/T_{\rm cav})$, $T_{\rm cav}=220$ $\mu$s, which could be expected on the basis of cavity relaxation parameters.
The red curve is the same Rabi oscillation but with stronger damping, $T_{\rm cav}=45$ $\mu$s. As we can see the red curve almost agrees with the data; the first two minima are somewhat lower than the data but the analysis does not include center-of-mass motion, finite duration of experiment, and the resulting frequency spread and detunings. 

So the red curve yields a reasonable agreement with experiment. The problem is that the cavity lifetime was 220 $\mu$s and not 45 $\mu$s, and the data should coincide with the blue curve.
This observation agrees with the remark of Brune {\it et al.\/} that `cavity relaxation plays a marginal role in the decrease of oscillation'. The authors further write: `Dark counts (...) are one of the main causes of oscillation damping(...). Decoherence by collisions with background gas may also contribute to the oscillation relaxation' 
(\cite{Haroche}, p. 1801). The relaxation observed experimentally thus appears to be stronger than expected, but the authors do not discuss the subtlety in much detail. 
In order to fit the data they used `damped sinusoids', but did not explain what kind of a damping factor (or factors?) were employed. 
Also the Fourier analysis was performed on time-symmetrized signals, and not just on damped ones. This is important since symmetrized signals are closer to  revivals-collapses typical of beats than to damped oscillations. In the light of the results we discuss in the present paper it is clear that a more detailed analysis of experimental damping factors is required.

In Fig.~3 we show the result with the same mixed initial condition but now computed by means of the reducible representation. The parameters are  $\Delta=0$, $g_{\rm ph}=47$ kHz, $Z=0.1$, and $N=280$. The blue curve involves no damping (ideal cavity); the red curve is additionally damped by $\exp(-t/T_{\rm cav})$, $T_{\rm cav}=220$ $\mu$s. The first two minima are again lower than the data, but the general agreement with experiment is acceptable. 
As one can see the damping is caused here mainly by the beats. Unfortunately we did not possess all the data needed for a realistic comparison with the experiment.

Now, can we distinguish between true damping and beats? In principle yes, but we must wait longer and have better cavities. Fig.~4 shows the dynamics of $w(t)$ in the reducible representation with exact vacuum initial condition, ideal cavity, $Z=0.1$, $N=280$, 
$\Delta=0$, and $g_{\rm ph}=47$ kHz for $0<t<0.0015$ s. Fig.~5 shows the same dynamics but now additionally damped by $\exp(-t/T_{\rm cav})$, $T_{\rm cav}=500$ $\mu$s. In principle, in a roughly twice better cavity we might see the revival. It must be stressed that this revival has no counterpart if field is quantized in irreducible representation, and vanishes if $N\to\infty$ (the limit taken with $Z$ fixed). Fig.~6 shows the dynamics of $w(t)$ in ideal cavity, exact vacuum, and for $N=10000$, the remaining parameters being kept as before. This case is interesting since up to 100 $\mu$s, that is the time available in the discussed experiment, the plot is practically indistinguishable from the result based on irreducible representations, whereas for longer times we observe collapse and revival. The greater $NZ$, the later comes the revival.

The examples show that the origin of relaxation should be carefully reexamined.

\subsection{General coherent-state initial condition}

Let us now replace vacuum by a general $|\uu z\rangle$ but for simplicity keep the initial atomic state to be $|+\rangle$. Then
\begin{widetext}
\be
w(t)
&=&
\langle+|R_z(t)|+\rangle
\\
&=&
\frac{1}{2}
-
\sum_{s=0}^N
\sum_{n=0}^\infty
\frac{|g_{\rm ph}|^2 (n+1)}{Z}\frac{s}{N}
\frac{\sin^2\big(t\sqrt{\Delta^2/4+|g_{\rm ph}|^2 (n+1)s/(ZN)} \big)}
{\Delta^2/4+|g_{\rm ph}|^2 (n+1)s/(ZN)}
\frac{|z\sqrt{\frac{s}{N}}|^{2n}}{n!}
e^{-|z\sqrt{\frac{s}{N}}|^2}
\left(
\begin{array}{c}
N\\s
\end{array}
\right)
Z_\omega^s (1-Z_\omega)^{N-s}\nonumber.
\ee
The limiting form, for $N\to\infty$, is again familiar
\be
\lim_{N\to\infty}w(t)
&=&
\frac{1}{2}
-
\sum_{n=0}^\infty
|g_{\rm ph}|^2 (n+1)\chi_\omega
\frac{\sin^2\big(t\sqrt{\Delta^2/4+|g_{\rm ph}|^2 (n+1)\chi_\omega} \big)}
{\Delta^2/4+|g_{\rm ph}|^2 (n+1)\chi_\omega}
\frac{|z_{\rm ph}\chi'_\omega|^{2n}}{n!}
e^{-|z_{\rm ph}\chi'_\omega|^2}.
\ee
\end{widetext}
For the same reasons as those discussed in the preceding subsections we obtain the standard formula but with the cut-offs $\chi_\omega=Z_\omega/Z$, 
$\chi'_\omega=\sqrt{\chi_\omega}$, and renormalized $e_{\rm ph}=e_0\sqrt{Z}$, 
$z_{\rm ph}=z\sqrt{Z}$. The rescaling $z\mapsto z_{\rm ph}$ is exactly analogous to 
(\ref{Z-eigen}), up to the presence of the cut-off, a negligible modification if $\chi_\omega\approx 1$. 

We again compare theory with the experiment of the Paris group. Fig.~7 shows the predictions based on the standard (irreducible) formalism and corresponding to Fig.~2B from \cite{Haroche}. We have the same problem as before: In order to fit the data we have to assume the cavity lifetime $T_{\rm cav}=50$ $\mu$s, which is much worse than in the experiment. On the contrary, the reducible formalism leads to a reasonably looking curve even 
for $T_{\rm cav}=220$ $\mu$s if we choose $NZ=42$ (Fig.~8). In these data we did not have access to the error bars. Fig.~9 and Fig.~10 show the plots corresponding to Fig.~2C in 
\cite{Haroche}. Finally, Fig.~11 illustrates the correspondence principle: Given any data collected in a finite time interval the reducible representation is capable of reconstructing the experimental points with arbitrary precision. 

\begin{figure}
\includegraphics{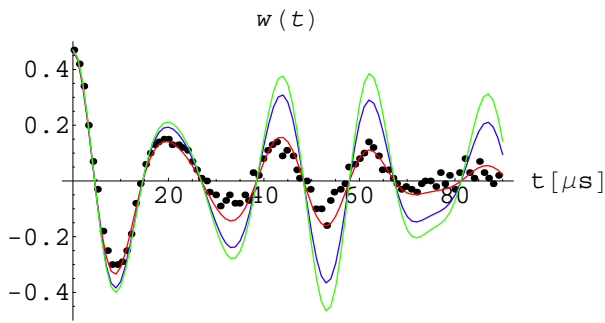}
    \caption{
    $w(t)$ computed on the basis of irreducible representations for a coherent state vs. experiment (Fig.~2B in \cite{Haroche}), with $z=\sqrt{0.4}$, $T=0$~K, $g_{\rm ph}=47$~kHz. Cavity parameters: $T_{\rm cav}=50$ $\mu$s (red), 
    $T_{\rm cav}=220$ $\mu$s (blue), and ideal cavity (green). The two level system is in a mixed state with $p_+=0.97$. The plot roughly coinciding with the data corresponds to a cavity that was more than four times worse than the one actually used in the experiment.
    }
\end{figure}
\begin{figure}
\includegraphics{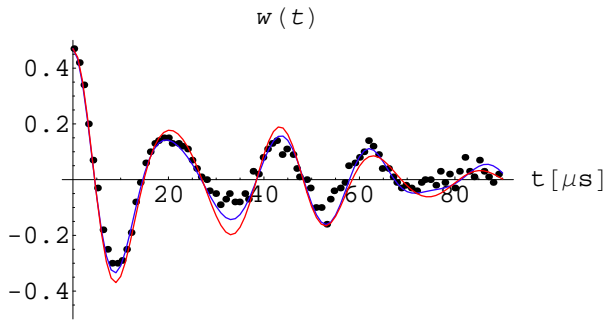}
    \caption{
    $w(t)$ (red) computed on the basis of the reducible representation, with $N=420$, $Z=0.1$, and $T_{\rm cav}=220$ $\mu$s, for the same coherent state as in Fig.~7. The blue line is the plot based on the standard formalism with $T_{\rm cav}=50$ $\mu$s. The reducible formalism produces the desired result with a realistic value of the damping parameter.
    }
\end{figure}
\begin{figure}
\includegraphics{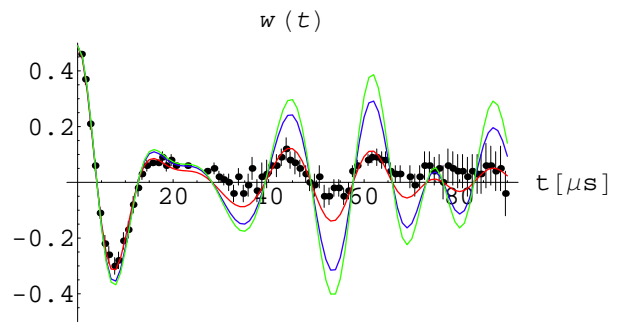}
    \caption{$w(t)$ computed on the basis of irreducible representations for a coherent state vs. experiment (Fig.~2C in \cite{Haroche}), with $z=\sqrt{0.85}$, $T=0$~K, $g_{\rm ph}=47$~kHz. Cavity parameters: (a) $T_{\rm cav}=50$ $\mu$s (red), (b) 
    $T_{\rm cav}=220$ $\mu$s (blue), and (c) ideal cavity (green). The two level system is in a mixed state with $p_+=0.99$. 
    }
\end{figure}
\begin{figure}
\includegraphics{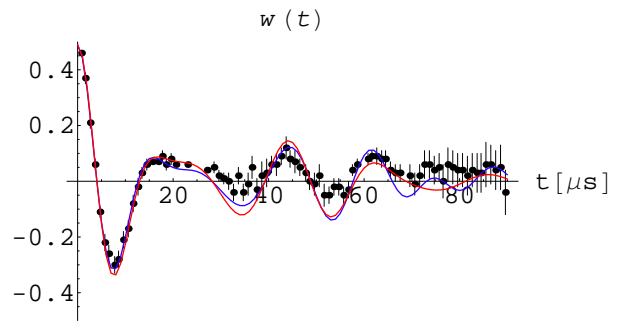}
    \caption{
    $w(t)$ (red) computed on the basis of the reducible representation, with $N=420$, $Z=0.1$, and $T_{\rm cav}=220$ $\mu$s, for the same coherent state as in Fig.~9. The blue line is the plot based on the standard formalism with $T_{\rm cav}=50$ $\mu$s.
    }
\end{figure}
\begin{figure}
\includegraphics{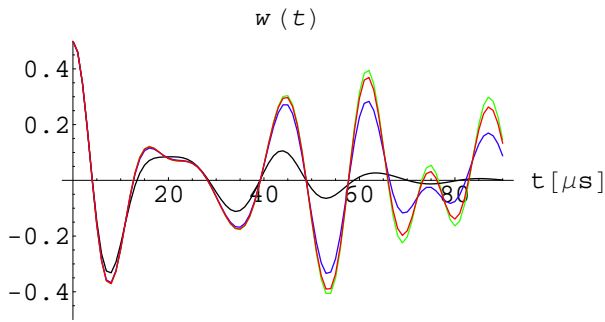}
    \caption{Correspondence principle in action. The green curve represents prediction of the standard theory for $w(t)$ with the coherent state initial condition, $z=\sqrt{0.85}$, 
    $p_+=1$, in ideal cavity and zero temperature. The remaining curves are the predictions based on reducible representations with $Z=0.1$, $z_{\rm ph}=\sqrt{0.85}$, for: $N=200$ 
    (black), $N=2000$ (blue), and $N=10000$ (red). 
    }
\end{figure}

\section{Conclusions}

The main conclusions are the following. First of all, the evolution of atomic inversion during a finite time interval can always be reconstructed by means of some $N<\infty$ representation. So, finite $N$ representation can be regarded as a generalization of standard quantum optics. This is a consequence of the fact that there exists the correspondence principle $N\to\infty$, analogous to $\hbar\to 0$ or $c\to\infty$. To our surprise at least some part of the available data seems to be more consistently explained with finite $N$ than with the standard formalism. In order to find the {\it proof\/} that $N<\infty$ is physical, we need to find a revival of a decaying vacuum Rabi oscillation.

\acknowledgments

This work is a part of the Polish Ministry of Scientific Research and Information Technology 
(solicited) project PZB-MIN 008/P03/2003.

\section*{Appendix A: `Reducible field quantization' in questions and answers}

Here we have collected the points we think are crucial for a correct understanding of the formalism based on $N<\infty$ representations.

\subsection{What if $Z_\omega=1$?}

In principle this case is not excluded. Then the vacuum is monochromatic, i.e. 
\be
|O\rangle &=& |\bm p,0\rangle,\\
|\uu O\rangle &=& |\bm p,0\rangle\otimes \dots\otimes |\bm p,0\rangle,
\ee
$Z=1$, and $\chi_\omega=1$ (exactly).
In such a vacuum a photon with momentum different from $\bm p$ cannot occur. 
In particular no resonance fluorescence is then possible.
The law of large numbers is then trivial since we are dealing with the Bernoulli process with probability of success equal to 1. The parameter $N$ then cancels in all the formulas for 
$w(t)$. However, this state is not very physical and there is no reason to believe such a vacuum can be encountered in experiment.

\subsection{What is the meaning of $\chi_\omega\approx 1$?}

What it means is that the probability $Z(\bm k)$, treated as a function 
$\bm k\mapsto Z(\bm k)$,  is very flat in some part of its domain  (termed in \cite{2,3,4} the {\it quantum optics regime\/}). Our resonant frequency is assumed to belong to the flat region of the distribution, i.e. $\chi(\bm p)$ is close (or even equal) to 1. 
The function $\chi(\bm k)=Z(\bm k)/Z$ has then all the features typical of the cut-off functions employed in quantum optics. However, $Z(\bm k)$ cannot be a constant function since then the wave function normalization condition $\sum_k Z(\bm k)=\sum_k 
|O(\bm k)|^2=1$ would not be fulfilled. The cut-off is a consequence of square-integrability of the wave function. Let us add that the wave function $O(\bm k)$ has a status analogous to the wave function of the Universe. 

\subsection{How small is $Z$?}

We do not know. However, if differences between $\chi(\bm k)$ and 1 are negligible in the quantum optics regime then many different momenta are equally probable and $Z$ (which is the maximum of the probability) must be a very small but nonzero number. Then, on the other hand, $N$ must be very large, and practically the parameter that controls finite $N$ representations is the product $NZ$ which, as we have seen, may be of the order of hundreds or thousands.
In a sense, the smaller $Z$ the better since then instead of two parameters, $N$ and $Z$, we have a single one: $NZ$. 

\subsection{What is the link between $N<\infty$ representations and oscillator wave packets?}

Take a nonrelativistic oscillator with the Hamiltonian
\be
H=
P^2/2m+m\Omega^2 Q^2/2
\ee
and assume that $\Omega$ is not a parameter but an operator which nevertheless commutes with 
$P$ and $Q$. This modification seems trivial, but is not if $\Omega$ is nontrivial. If $\Omega=\sum_\omega \omega I_\omega$ is its spectral representation with spectral projectors $I_\omega$, then $[I_\omega,P]=[I_\omega,Q]=[I_\omega,H]=0$. Define $H_\omega=I_\omega H$. Then
\be
H &=& \sum_\omega H_\omega,\\
H_\omega &=& \frac{\hbar\omega}{2}\big(a_\omega^{\dag}a_\omega+a_\omega a_\omega^{\dag}\big)
\nonumber\\
&=&\hbar\omega a_\omega^{\dag}a_\omega+\frac{\hbar\omega}{2}I_\omega,\\
{[a_\omega,a_{\omega'}^{\dag}]} &=& \delta_{\omega\omega'}I_\omega,\\
\sum_\omega I_\omega &=& I.
\ee
This is precisely an example of $N=1$ representation. Taking $N$ such (noninteracting!) oscillators we arrive at  a $N<\infty$ representation. The oscillator can exist in superposition of different eigenstates of $\Omega$. A gas consisting of $N$ such oscillators, all in identical ground-state wavepackets, is a Bose-Einstein condensate. But this is simultaneously our vacuum state. 

\subsection{What about Lorentz invariance?}

No problem. A relativistic formalism is developed in \cite{2,3,4} (although \cite{2} is yet preliminary --- one should work with \cite{4}). We plan to redo the Jaynes-Cummings calculations in the representation of \cite{4}.

\subsection{Are there modifications of the blackbody radiation?}

Not really. The analysis of the problem given in \cite{1} was premature since the role of the large-$N$ limit and the law of large numbers was not yet understood in this context. A single oscillator, i.e. the 
$N=1$ representation, has spectrum  {\it identical\/} to the standard one (i.e. $E_n=\hbar\omega(n+1/2)$; the difference is that here $\omega$ is also an eigenvalue and not a parameter). The field is a gas of $N$ such oscillators. The only subtlety is that the gas is finite and, as shown in \cite{4}, the statistics may be based on R\'enyi $\alpha$-entropies with $\alpha=1-1/N$. The limit $N\to\infty$ is then equivalent to $\alpha\to 1$, but this is the Shannon limit of $\alpha$-entropies. 

\subsection{What about the divergences?}

This is not yet completely clear, but all the examples discussed so far show that the cut-offs occur in the correct places. The work on full QED, loop integrals included, is in progress. 
There is one element that is not completely controlled yet. Namely, if we work in full space and not in a cavity (such as the classical current example discussed in \cite{4}) then it is natural to start with $N=1$ representations that involve spectral projectors $I_\omega$ corresponding to plane waves (i.e. these are not really projectors because of Dirac delta normalizations). Such a procedure is useful in some cases, but the final formalism should always use actual projectors in order to avoid introducing artificial infinities coming from the terms $I_\omega^2$. The `modes' should be associated with basis vectors in a Hilbert space, and not with plane waves. A trivial way out is to work always with finite volumes, but the formalism then lacks elegance. Some polishing of the formalism is here yet needed.

\subsection{What about vacuum energy?}

For a finite $N$ the vacuum part of the Hamiltonian is a well defined operator from the center of the algebra (i.e. commutes with everything). We can remove it by a unitary transformation which is well defined (this is what we implicitly do in the present paper, and sometimes refer to the procedure as a {\it vacuum picture\/} \cite{2}). The average energy of a single oscillator is finite by assumption (this is a condition on the domain of the $N=1$ Hamiltonian and means that 
$\sum_\omega \omega Z_\omega<\infty$). For an arbitrary $N$ the vacuum energy is $N$ times the average energy of the $N=1$ case and, of course, diverges with $N\to\infty$. However, in this sense the mass of a glass of water diverges if one treats the thermodynamic limit too literally. By the way, the vacuum energy of Dirac electrons, as discussed in \cite{3}, is {\it negative\/}. The discussion of vacuum energy in QED must involve both fermions and bosons, and then we have a difference of two finite expressions which may be well defined even in the limit of large $N$. 

\subsection{Isn't what we do a cut-off regularization in disguise?}

No, because there are no cut-offs in operators, and in the Hamiltonians in particular. It is true that in effect the end result is similar, especially in the limit $N\to\infty$. 
But since there is no cut-off in the Hamiltonian, there cannot be any cut-off dependence in its spectrum! This point is very important since, in principle, it can lead to yet another direct test of the $N<\infty$ representation. In a forthcoming paper we shall discuss the spectral properties of the Jaynes-Cummings model. The experimental aspects have been worked out by the Caltech group \cite{Kimble1,Kimble2}.

\section*{Appendix B: Technicalities}

\subsection{Spectral projectors of $a(s)^{\dag}a(s)$ vs. $a(s)$ and $a(s)^{\dag}$}

For $s=0$ we find $a(s)=0$. So consider $s>0$, $m>0$,  and
\be
\Pi(m,s)a(s) \Pi(n,s)
&=&
\frac{N}{ms}\Pi(m,s)a(s)^{\dag}a(s)a(s) \Pi(n,s)\nonumber\\
&=&
\frac{n-1}{m}\Pi(m,s)a(s)\Pi(n,s)\nonumber.
\ee
Hence
$
\frac{n-1-m}{m}\Pi(m,s)a(s)\Pi(n,s)
=
0
$
and for $0<m\neq n-1$ one finds 
\be
\Pi(m,s)a(s)\Pi(n,s)
&=&
0\nonumber
\ee
and, by Hermitian conjugation,
\be
\Pi(n,s)a(s)^{\dag}\Pi(m,s)
&=&
0.\nonumber
\ee
Analogously, for $m=0$, $n>0$, 
\be
\Pi(0,s)a(s) \Pi(n,s)
&=&
\frac{1}{n}\Pi(0,s)a(s)\Pi(n,s)
\nonumber
\ee
and
\be
\frac{n-1}{n}\Pi(0,s)a(s)\Pi(n,s)=0.
\nonumber
\ee
Now assume 
$a(s)\Pi(0,s)\neq 0$. Then there exists a vector $|\psi\rangle$ such that 
$a(s)\Pi(0,s)|\psi\rangle=:|\Psi\rangle\neq 0$.
However, 
\be
\langle \Psi|\Psi\rangle
&=&
\langle \psi|\Pi(0,s)a(s)^{\dag}a(s)\Pi(0,s)|\psi\rangle=0\nonumber.
\ee
Contradiction. Therefore $a(s)\Pi(0,s)= 0$. 

\subsection{Spectral projectors of $X(s)$ vs. $\Omega(s)$}

For $n>0$
\be
{[R_+a(s),\hat \Pi(n,s)]}
&=&
R_+\big(a(s)\Pi(n,s)-\Pi(n-1,s)a(s)\big).\nonumber
\ee
Employing the formulas from the previous subsection we find
\be
a(s)\Pi(n,s) &=& \Pi(n-1,s)a(s)\Pi(n,s),\nonumber\\
\Pi(n-1,s)a(s) &=& \Pi(n-1,s)a(s)\Pi(n,s),\nonumber
\ee
and 
$[R_+a(s),\hat \Pi(n,s)]=0$. By Hermitian conjugation $[R_-a(s)^{\dag},\hat \Pi(n,s)]=0$. 
For $n=0$ 
\be
{[R_+a(s),\hat \Pi(0,s)]}
&=&
R_+a(s)\Pi(0,s)=0.\nonumber
\ee
By conjugation $[R_-a(s)^{\dag},\hat \Pi(0,s)]=0$ and, for all $n\geq 0$, 
$[\Omega(s),\hat \Pi(n,s)]=0$. 

\subsection{Average $\langle \uu z|\Pi(n,s)|\uu z\rangle$}

By construction $a(s)|\uu O\rangle=0$. 
Introducing the rescaled operators $\tilde a(s)=a(s)/\sqrt{s/N}$, satisfying 
$[\tilde a(s),\tilde a(s)^{\dag}]=P(s/N)$, we check that
\be
{}&{}&a(s)^{\dag}a(s)\big(a(s)^{\dag}\big)^n |\uu O\rangle
=
\Big(\frac{s}{N}\Big)^{1+\frac{n}{2}}
\tilde a(s)^{\dag}\tilde a(s)\big(\tilde a(s)^{\dag}\big)^n |\uu O\rangle
\nonumber\\
&{}&\pp=
=
\Big(\frac{s}{N}\Big)^{1+\frac{n}{2}}
n\big(\tilde a(s)^{\dag}\big)^n |\uu O\rangle
=
\frac{s}{N}
n\big(a(s)^{\dag}\big)^n |\uu O\rangle.
\nonumber
\ee
Therefore 
\be
\Pi(n',s)\big(a(s)^{\dag}\big)^n |\uu O\rangle
&=&
\delta_{n,n'}\big(a(s)^{\dag}\big)^n |\uu O\rangle,
\nonumber\\
\Pi(n,s)|\uu z\rangle
&=&
\frac{z^n}{n!}
e^{-\frac{1}{2}|z|^2 (s/N)}
\big(a(s)^\dagger\big)^n
|\underline{O}\rangle,\nonumber
\ee
and 
\be
{}&{}&\langle \uu z|\Pi(n,s)|\uu z\rangle
=
(s/N)^n\frac{|z|^{2n}}{n!}
e^{-|z|^2 (s/N)}
\langle \underline{O}|P(s/N)|\underline{O}\rangle\nonumber\\
&{}&\pp=
=
\frac{|z\sqrt{s/N}|^{2n}}{n!}
e^{-|z\sqrt{s/N}|^2}
\left(
\begin{array}{c}
N\\
s
\end{array}
\right)
Z_\omega^s (1-Z_\omega)^{N-s}.\nonumber
\ee
\begin{widetext}
\subsection{Explicit form of $R_3(t)$}

We have to simplify
\be
{}&{}&
R_3(t)\label{eq27}
\\
&{}&\pp=
=
\bigg[
\cos\big(\sqrt{\Delta^2/4+|g|^2 {X}}t\big)
+
i
\frac{\sin\big(\sqrt{\Delta^2/4+|g|^2 {X}}t\big)}{\sqrt{\Delta^2/4+|g|^2
{X}}}\Omega
\bigg]
R_3
\bigg[
\cos\big(\sqrt{\Delta^2/4+|g|^2 {X}}t\big)
-
i
\frac{\sin\big(\sqrt{\Delta^2/4+|g|^2 {X}}t\big)}{\sqrt{\Delta^2/4+|g|^2 {X}}}
\Omega
\bigg].
\nonumber
\ee
\end{widetext}
One checks that
$[X,R_3]=[X, \Omega]=
[X, R_+ a_\omega]
=[X, R_- a^*_\omega]
=[X, a^*_\omega a_\omega]
=0$,
$
[\Omega, R_3]
=
\bar g R_- a^*_\omega
-
g R_+ a_\omega
$,
$$
\Omega R_3\Omega
=
\big(\Delta^2/4-|g|^2 X\big) R_3
+
\frac{\Delta}{2}\big(
g R_+ a_\omega
+
\bar g R_-a^*_\omega\big).
$$
Employing these formulas we arrive at (\ref{R3tgen}).

\end{document}